\documentclass[epjCONF,columns]{svjour} 
\usepackage{amssymb,amsmath}
\usepackage{graphics}
\usepackage[varg]{txfonts} % Times fonts
\usepackage[latin1]{inputenc}
\session-title{Hadron Collider Physics Symposium 2011}
\begin{document}
\title{Photon polarisation in $b \rightarrow s\gamma$ using $B \rightarrow$ K$^*$e$^+$e$^-$ at LHCb }
\author{Michelle Nicol (on behalf of the LHCb collaboration)\inst{1}\fnmsep\thanks{\email{nicol@lal.in2p3.fr}} }
\institute{Laboratoire de l'Acc\'{e}l\'{e}rateur Lin\'{e}aire, Orsay, France}
\abstract{The $b \rightarrow s\gamma$ transition proceeds through flavour changing neutral currents, and thus is particularly sensitive to the effects of new physics. An overview of the method to measure the photon polarisation at the LHCb experiment via an angular analysis of $B \rightarrow K^*e^+e^-$ at low $q^2$ is presented. The status of the $B \rightarrow K^*\mu^+\mu^-$ analysis with 309 pb$^{-1}$ of $pp$ collisions at $\sqrt{s}$=7 TeV at LHCb is also given.} %end of abstract
\maketitle
\section{Introduction}
\label{intro}
Although the branching ratio of $b \rightarrow s\gamma$ has been measured to be consistent with Standard Model (SM) predictions, new physics could still be present and detectable through the analysis of  details of the decay process. In particular, the photon from the b is predominantly left handed in the SM, whereas additional right handed currents can arise in certain new physics models, such as the Left-Right symmetric models, or in some supersymmetric models \cite{nonSM}. Access to the polarisation information is available via an angular analysis of $B \rightarrow K^*e^+e^-$.

Hadronic form factors render theoretical  prediction over the whole $q^2$ (the dilepton invariant mass squared) range difficult. However, it has been shown that these uncertainties are controllable at low $q^2$, where the photon term dominates, and certain asymmetries providing information on the photon polarisation can be formed. \cite{RefJ}
\section{$B \rightarrow K^*\mu^+\mu^-$ status at LHCb}
\label{status} With 309 pb$^{-1}$ of $pp$ collisions at $\sqrt{s}$=7 TeV, collected in three months during the first half of 2011, the forward backward asymmetry of the dilepton system, $A_{\normalfont FB}$ has been measured \cite{muon} using $B \rightarrow K^*\mu^+\mu^-$ events, (as is shown in Fig. \ref{afb}), along with $F_L$, the K$^{*}$ longitudinal polarisation (Fig. \ref{fl}); an input required for the photon polarisation measurement. These observables have been measured as being in good agreement with SM predictions, \cite{SM}, implying a SM like Wilson Coefficient $C_7$,  but still allowing for the existence of $C_{7}^{'}$ (right handed currents).
As stressed above, the measurement is most sensitive at low $q^2$. It would therefore be preferable to perform the analysis using electrons. However, experimentally it is more challenging to observe electrons than muons, primarily due to the fact that muons provide a very clean signature to trigger on. With 309 pb$^{\normalfont-1}$ of LHCb data, $B \rightarrow K^*\mu^+\mu^-$ in the $q^2$ range 0-2 GeV has been observed, as is shown, along with other $q^2$ ranges, in Fig. {\ref{muon}. With the rest of the 2011 data, one can expect to see a $B \rightarrow K^*e^+e^-$ signal.
\begin{figure}
\resizebox{0.75\columnwidth}{!}{%
\includegraphics{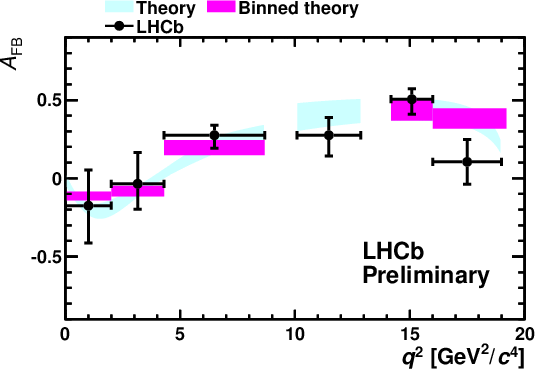} }
\label{afb}
\resizebox{0.75\columnwidth}{!}{%
\includegraphics{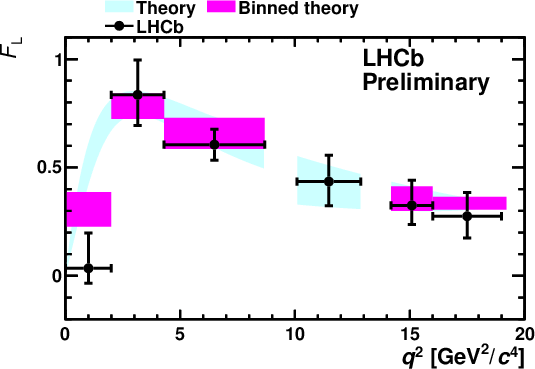} }
\label{fl}  
\caption{$A_{\normalfont FB}$ and F$_{\normalfont L}$ as a function of $q^2$, as measured at LHCb with $B \rightarrow K^*\mu^+\mu^-$ \cite{muon}. The SM predictions are given by the cyan (light) band, and this prediction integrated in the $q^2$ bins is indicated by the purple (dark) regions.}
\end{figure}
\begin{figure}
\resizebox{0.95\columnwidth}{!}{%
\includegraphics{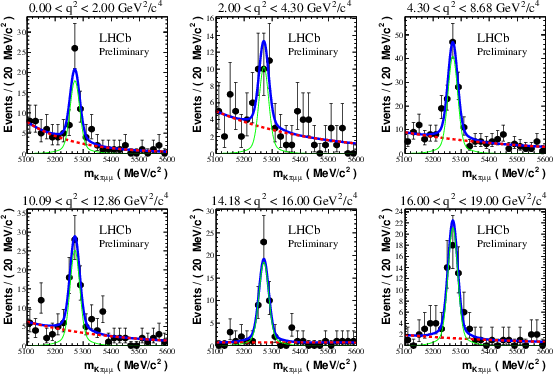} }
\caption{The mass distributions of $B \rightarrow K^*\mu^+\mu^-$ in six $q^2$ bins. The solid line shows
a fit with a double-Gaussian signal component (thin-green line) and an exponential background component (dashed-red line).}
\label{muon}       
\end{figure}
\begin{figure}
\resizebox{0.75\columnwidth}{!}{%
\includegraphics{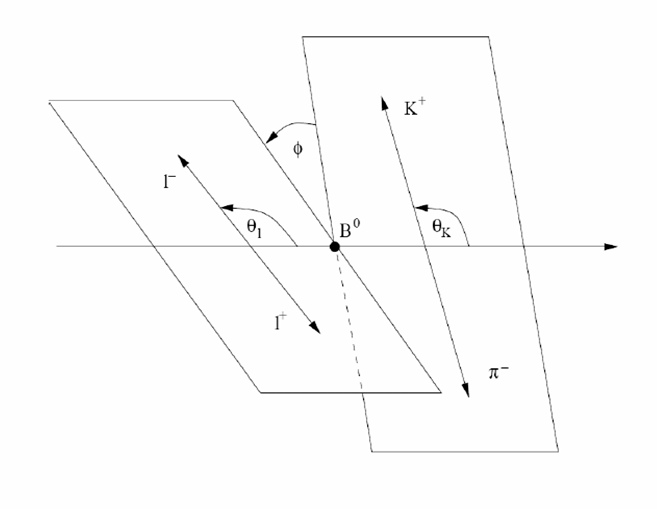} }
\caption{Definition of the angles $\phi$, $\theta_K$ and $\theta_L$ in the decay $B \rightarrow K^*e^+e^-$.}
\label{angles}       
\end{figure}
\section{Analysis formalism}
\label{formalism}
$B \rightarrow K^*e^+e^-$ can be uniquely described by four variables: $q^2$ and three angular variables, $\theta_L$, $\theta_K$ and $\phi$, (the definitions of which can be seen in Fig. \ref{angles}). Following the formalism as described in \cite{krug}, the differential decay distribution can be written in terms of these variables as:
\begin{align}
\label{formal}
&\frac{d\Gamma}{dq^2d\cos\Theta_ld\cos\Theta_Kd\phi} =\notag\\ 
&\frac{9}{32\pi}[I_1\left(\cos\Theta_K\right)+ I_2\left(\cos\Theta_K\right)\cos2\Theta_l + \notag \\
&I_3\left(\cos\Theta_K\right)\sin^2\Theta_l\cos2\phi + I_4  \left(\cos\Theta_K\right)\sin2\Theta_l\cos\phi +\notag\\
&I_5\left(\cos\Theta_K\right)\sin\Theta_l\cos\phi + I_6\left(\cos\Theta_K\right)\cos\Theta_l + \notag\\
&I_7\left(\cos\Theta_K\right)\sin\Theta_l\sin\phi  + I_8\left(\cos\Theta_K\right)\sin2\Theta_l\sin\phi\ + &\notag\\
&I_9\left(\cos\Theta_K\right)\sin^2\Theta_l\sin2\phi]
\end{align}
When measuring this rate at LHCb, the 3D angular acceptance, $\varepsilon\left(\cos\Theta_l,\cos\Theta_K,\phi\right)$ must also be taken into account. It is assumed to be factorisable as the products of $\varepsilon_1$, the acceptance as a function of $\phi$, and $\varepsilon_D$, the acceptance as a function of $\cos\Theta_K$ and $\cos\Theta_L$. Furthermore, assuming that $\varepsilon_1$ is an even function, Equation \ref{formal} can be simplified by performing the $\phi$ transformation that if $\phi$ $>$0, then $\phi$=$\phi$+$\pi$. A similar transformation can be performed for $\cos\Theta_L$. Equation \ref{formal} can then be written as:
\begin{align}
\label{formal2}
&\frac{d\Gamma}{dq^2d\cos\Theta_ld\cos\Theta_Kd\phi} =\notag\\ 
&\frac{9}{32\pi}[I_1\left(\cos\Theta_K\right)+ I_2\left(\cos\Theta_K\right)\cos2\Theta_l + \notag \\
&I_3\left(\cos\Theta_K\right)\sin^2\Theta_l\cos2\phi + I_9\left(\cos\Theta_K\right)\sin^2\Theta_l\sin2\phi] \notag \\
&\times \varepsilon_D\left(\cos\Theta_l,\cos\Theta_K\right)
\end{align}
In order to minimize theoretical uncertainties, it is desirable to measure ratios of the amplitudes. Neglecting the lepton mass, the remaining I terms in equation \ref{formal2} can be written in terms of three such parameters, $\mathrm{F_L, A_T^{(2)},A_{Im}}$:
\begin{equation}
\begin{split}
\mathrm{F_L} &= \frac{\mathrm{\left|A_0\right|^2}}{\mathrm{\left|A_0\right|^2}+ \left|A_\bot\right|^2 + \left|A_\|\right|^2}\\
\mathrm{A_T^{(2)}} &= \frac{\mathrm{\left|A_\bot\right|^2 - \left|A_\|\right|^2}}{\mathrm{\left|A_\bot\right|^2 + \left|A_\|\right|^2}}\\
\mathrm{A_{Im}} &= \frac{\mathrm{\Im(A^*_{\bot L}A_{\bot L})-\Im(A^*_{\bot R}A_{\bot R})}}{\mathrm{\left|A_0\right|^2}+ \left|A_\bot\right|^2 + \left|A_\|\right|^2}
\end{split}
\end{equation}
When expressed in terms of the helicity amplitudes, for small real values of $\frac{A_R}{A_L}$, one obtains $A_{T}^{(2)}\approx-2\frac{A_{right}}{A_{left}}$.

\section{$B\rightarrow$ K$^*$e$^+$e$^-$ Monte Carlo studies}
\label{MC}
Although work is ongoing on the analysis of the $B\rightarrow$ K$^*$e$^+$e$^-$ data, and yield predictions from Monte Carlo (MC) have been validated using the control channel $B\rightarrow$ K$^*$J$/\Psi$ with J$/\Psi \rightarrow$(e$^+$e$^-$), there is not yet, at the time of this conference, enough data to perform the analysis or test the fitting procedure. Toy 
MC studies have therefore been carried out for this purpose \cite{schune}.
190k signal events were generated using EvtGen, and separated into files containing 250 events: the predicted yields  from MC studies with 2fb$^{\normalfont-1}$ at a centre of mass energy of 14 TeV, excluding effects from LHCb's high level trigger. By performing the fit on each file, it is shown that with 200-250 signal events and a signal to background ratio of the order of 1, a precision of 0.2 is attainable on $\mathrm{A_T^2}$, equivalent to an accuracy on the fraction of wrongly polarised photons of 0.1. An example of one of the fits for one toy MC study can be seen in Fig. \ref{fit}. The analysis also demonstrates that the measurements are not sensitive to the knowledge of the angular acceptance, and hence shall not be systematics limited. 
\begin{figure}[h!]
\resizebox{\columnwidth}{!}{%
\includegraphics{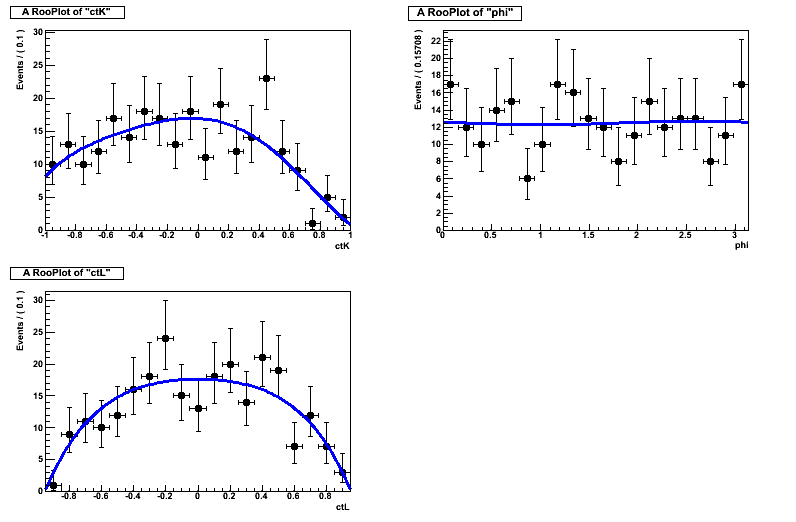} }
\caption{Example of the fit of $\cos\Theta_L$, $\cos\Theta_K$ and $\phi$ for one toy  MC study containing 250 signal events and no background events.}
\label{fit}       
\end{figure}

\end{document}